# A New Innovation Concept on End user's Contextual and Behavioral Perspectives

Reem Aman[1] and Shah J. Miah[2] and Janet Dzator[3]

[1] Ministry of Education, Riyadh, Kingdom of Saudi Arabia
Newcastle Business School, University of Newcastle, NSW, Australia
reem.aman@uon.edu.au
[2] Newcastle Business School, University of Newcastle, Newcastle NSW, Australia
shah.miah@newcastle.edu.au
[3] Newcastle Business School, University of Newcastle, Newcastle NSW, Australia
janet.dzator@newcastle.edu.au

**Abstract.** This research study explores this original phenomenon for proposing a new concept that will act as an overarching descriptor of innovation types: idea, object, and behavior. This proposed concept, relating to intangible innovation, will explain the sequence within one or many connected intangible activities that provide novelty to its end-user relative to previous activities and practices. Using a design science research approach, the study comprises two goals: a) identifying opportunities and issues to measure intangible inputs to the innovation and b) proposing a framework for extending the existing innovation theories that to better capture intangible end-user innovation and its diffusion insights in their online environment across nations.

**Keywords:** diffusion of innovation theory, user innovation, visual analytics, big data, design science research

## 1    Introduction

This study looks at an alternative paradigm of ICT application development for end user in their own context of innovation. We look at the purpose of technology to generate knowledge from information relevant to innovation support. This innovation focuses on informed actions by concentrating on method and task [1]. The entire design and inductive tasks would be considering as an exploratory and end user-oriented design science research that highlight end user's particular contextual situation approach for designing conceptual solution framework [2-4].

Innovation by individual end users in the household sector (HHS) is a widespread phenomenon on a national scale. Individual end users in the household sector modify and develop tangible and intangible innovations in household projects context. There are multiple conceptualizations for intangible innovations, which indicate the importance of conceptual clarity to distinguish between tangible and intangible innovations. It is important for research to develop conceptual clarity that could extend academic knowledge on understanding the phenomena, the nature of it as an individual project instead of idea, object or practice, its diffusion, and the impact of it on a national scale. Therefore, the exploration of studying this phenomenon will have theoretical and empirical implication as academic research.

Utilizing the end-user-oriented design understanding, this study proposes an over-arching conceptualization of innovation types by end users in the household sector, to illustrate and communicate the complexity of intangible innovation. It will explore end user innovation scope [5-7], through visual analytics of big data to discuss a new type of intangible innovation, scope innovation within a household project context. This conceptual framework will give a design basis of a fully functional end-user innovation support solution that provides decision support in relation to different intangible innovations in household sector.

## 2 Research Background

### 2.1 Intangible innovation

Intangible innovations can be difficult to distinguish, especially when it comes to its diffusion in end users' context. Therefore, it was suggested that operating innovation at a process level will assist it's identification and capturing its diffusion [8]. This can be challenging in a survey study, or impossible. Although, it is important to understand intangible innovation by end users in the household sector, It was argued that traditional methods might not be reliable [9]. For example, the lack of understanding by the individual of the nature of innovation, especially in a household sector setting, because they might associate it with a daily practice [9].

### 2.2 Intangible innovation by HHS

There are multiple conceptualizations for intangible innovations, which indicate the importance of conceptual clarity to distinguish between tangible and intangible innovations. This would be of paramount task to assist the classification and the measurement of individual end users' innovation impact in the household sector. Innovation by end users in the household sector received wide attention in the literature, since it was first captured in a national survey, which proposed it scale and scope [10]. End user innovation in the household sector is the modification or the development of innovation by individuals who are not commercial groups [10].

Multiple national survey studies illustrated the phenomena in several developed and developing countries [10-18]. Furthermore, it was argued that there are value propositions of tangible and intangible types of innovation developed and modified by individual end users.

### 2.2.1 Technique and service innovation

Technique innovation by end users was identified in an adventure sports context. It included novel movements in the water, physical movements, and environmental movements [19]. Technique innovation by individual end users was defined as a "new way of doing", done by an individual, as a skillful and planned activity, either involve a physical object, or operates in an environment with physical objects [20]. In the banking sector it was found that innovation in service was first initiated by end users from the household sectors, in developed [21] and developing countries [22]. Service innovation by end users is to meet self needs [21], and not for third party like in commercial settings [23].

### 2.2.2 Behavioral innovation

Another important dimension of end user innovation is the behavioral innovation which can be seen as a combined form of technique and service innovation. Behavioral innovation was defined as "one or a connected sequence of intangible problem-solving activities that provide a functionally novel benefit to its user developer relative to previous practice" [24]. However, a particular focus of this innovation is limited to develop systematic procedure for further guide and assistance.

This study will propose an overarching conceptualization of behavioral innovation, to illustrate and communicate the complexity of intangible innovation, when discussed in future research. The proposed conceptualization in this study will discuss a new type of intangible behavioral innovation, scope innovation, which include innovation types: idea, object, and practice. Also, it will act in contrast with tangible innovation. This contribution to the literature will need the development of a new design science model, to facilitate the study in capturing the scope and significant of intangible innovation by individuals in the household sector on a national scale, in detail.

### 2.2.3 HHS innovation scope

The exploration of scope innovation will add to our knowledge about end user innovation and extend the adopters categories in the diffusion of innovation theory [24]. For example, the topics that were discussed in relation to individual innovators characteristics include gender [25], personality [26], and capabilities [12]. The pilot study will enable the inclusion of additional variables that will be more relevant to characterized innovators based on extend of novelty. This will create a new set of innovators/adopter's categories that will extend the diffusion of innovation theory. Therefore, it is proposed in this study that the identification of tangible and intangible innovation diffusion will aid the exploration of innovators novelty based on the scope of their process innovation in a project context. Therefore, we develop the research question as: What attributes of the individual innovation are vital for facilitating diffusion and adoption of technological artefact?

## 3 Research Methodology

Improvement of the artefact design knowledge is an essential component of design science research [27]. Design research "…seeks to create innovations that define the ideas, practices, technical capabilities, and products through which the analysis, design, implementation, management, and use of information systems can be effectively and efficiently accomplished" [27, p. 76]. Therefore, this paradigm, going beyond the traditional qualitative and quantitative methods, provides supportive knowledge and guide for the artefact design (in this study, the artefact is the conceptual framework for scope innovation).

Design science also goes beyond the traditional sequential phase or step-oriented design methods ensuring that the problem-solving method is to be designed is fully problem informed, and enables creation of new generalizable problem-solving knowledge (e.g. prescriptive solution (design) knowledge) for similar issues [27, 28, 29]. Design studies have introduced various approaches such as action design research - ADR [30]; design science research methodology [31] and participatory action design research [32], although there are some limitations on each of them. Analyzing different approaches, we found that [31] six activity-oriented approach that has been somehow driven by the seven design guidelines of [27] together provide us a suitable synergy for supporting the objectives of our study. We therefore rebuild a suitable design approach that meets our goals both in articulating end user innovation realities and designing a problem-solving method that may integrate computational techniques. A conceptual framework is defined as a design research artefact that provides specifications of ways of performing tasks as well as for illustrating relationship between components (attributes, causes or factors that creates effect) within the particular problem context.

Different variables and the assumed relationships between those variables are included in the framework model and reflect the innovation expectation's goal directed activities in order to solve or address particular problem called wicked problems (e.g., comprises technical, human and organizational elements in the problems [27]. In our study, our approach is based on three phases (illustrated in Table. 1):

**Table 1.** Three design phases for conducting the study

| Phases | Tasks in our study |
|---|---|
| Design Phase 1: Problem realization and artefact types | Literature review; data analysis; gap analysis; end users' provisions; context analysis; problem articulation; artefact type selection |
| Design Phase 2: Artefact creation and evaluation, and | Identifying components of framework; Framework design; framework validation; framework justification and reformation parameter identification |
| Design phase 3: Research contributions of the artefact and communication of results | Dissemination of design ideas; design artefact and identification of generalizable parameter of the proposed framework; applicability analysis |

The three design phases for conducting the study are informed through [27] and articulated on [31] six activities. [27] suggested that design science research must talk about the creation of an innovation and purposeful development that may capture the problem situation, reality, and the key demands of the purpose in a specific problem domain. This implies that a collection of innovative conceptual artefact that can reinforce quality by creating effective design to meet the needs of the end users as well as being able to fulfil the process, users' and situational requirements within a problem space (e.g. household project contexts). The definitions of [27] establish two useful views that can help define a purposeful artefact design and its properties.

## 4    Theoretical framework

In the diffusion of innovation theory (DOI), diffusion is the spreading of properties by the penetration of invention through a process of imitation, and the properties resulting from invention, transmit from an individual to another individual [33]. Therefore, diffusion is a phenomenon operating on a micro-level. The essence of imitation between individuals in the diffusion process is a repetition that is conditioned by universal laws [33]. There are multiple models of innovation development [34]. Some of innovation development models that include invention and diffusion were based on the diffusion of innovation theory [35]. These models were used by economics and managers [36]. However, these innovations are bounded by market failure [14].

Innovations by end users diffuse in two forms C2B or C2C (consumer to businesses or consumer to consumer) [37]. In the first path the innovation is diffused to the market through firm's adoption. In the second path the innovation is either diffused through a community and into the market or be adopted by peers. In this path, the innovation ends up creating a new market [38]. In individual's innovation projects, end users develop innovation in a collaborative mode and in an individual mode [9]. In a collaborative mode, the innovation project has a distributed nature [39]. The end user re-innovates and diffuse the innovation within a community, where another member may require supportive information to adopt or rectify the innovation, which will then diffuse to another member for modification in their own context. This complex phenomenon is represented in open-source software development [39], but appropriate support framework are yet to be developed for different stakeholders in this sector.

## 5    Proposed conceptualization

Our proposed conceptual framework embraces components of intangible innovation beyond technique, service, or behavior as a scope innovation. These can connect parts of most theoretical and empirical studies in this problem domain on an individual level; ambiguously. End users' scope is one of the dimensions of process innovation diffusion and how it would be autonomous and systemically nurtured for end user enhancement. The proposed framework could provide supportive environment for the nature of innovation outcome. Figure 1 illustrates some initial components of the framework that will be a primary basis of our design study:

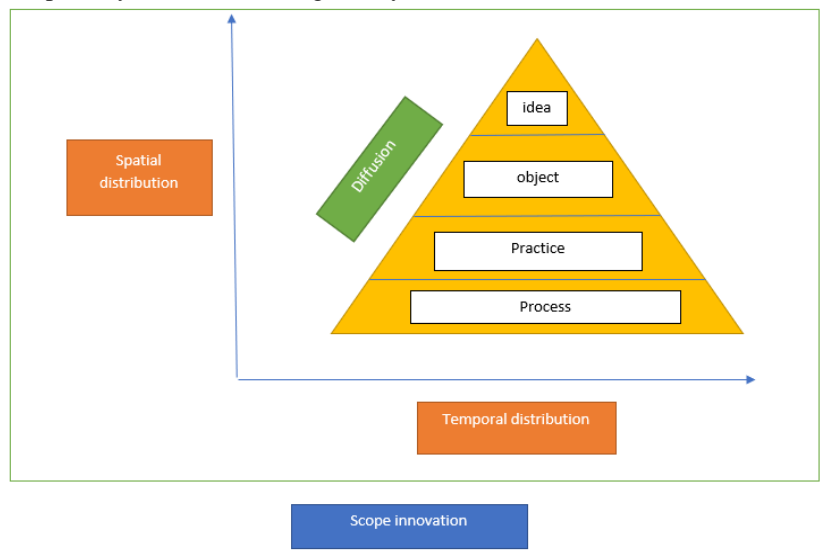

**Fig. 1.** Key components of the proposed end user innovation framework

In the proposed framework, we revised the scope was defined as "the extent of the area or subject matter that something deals with or to which it is relevant" that can be viewed as an opportunity. Furthermore, a scope can take the form of an investigation or evaluation [40] and defined in organizational development as the number of units associated with the innovation through diffusion and change [41]. For example, the scope of user innovation was discussed in a process innovation context and its magnitude was empirical demonstrated nationally [5] In the household sector, the view of individual end user innovation phenomena [42] shows promises that was firstly identified in modifications done by individuals. Also, it was argued that most individual end users modify or develop process equipment and software to meet their needs [43], and the two properties of scope: autonomous and dichotomous [41], which are used as a fundamental benchmarking of the entire framework design.

## 6 Discussion and conclusion

The study was to describe a new methodological foundation for designing an end user innovation design framework in the household sector. Design science research becomes the central element that would provide systematic guide and assistance for the entire design study in capturing end user innovation contextual details and converting them into a purposeful artefact (e.g. regarding intangible innovation). The three phases in the design methodology will enable us in designing relatively new concept for promoting intangible components of innovation that offers both further reinvention and invention support to end users. We deployed an openly online environment within peer-to-peer communication setting, to capture intangible and tangible innovation diffusion. This method will not replace traditional methods such as surveys, it will act as a complement to it. The filtering process of individual scope innovation will be captured to develop meta data of innovation types, innovation needs, nature of novelty, diffusion activities, and innovators characteristics, which will be initially based on secondary data. Our extended understanding would contribute to rectify the current form of DOI theory, including new components in relation to individual innovation needs in their own context.

This study recreated an alternative paradigm of ICT application development for end user in their own context of innovation. We redefined the context as "end-user own context of design innovation" in this paper for the first time in the literature. We believe that the purpose of technology is to assist end users for intervention support and produce new knowledge for innovation support. It will focus on informed actions by focusing on method and task [1]. The entire design and inductive tasks would be considering as an exploratory and end user-oriented design science research that highlight end user's particular contextual situation approach for designing conceptual solution framework [2, 3].

The literature of individual innovation in the household sector is at its emergent stage at a national scale. New understanding and knowledge should be reproduced in this sub-domain for new researchers and industry practitioners for more provisions of modifying and developing tangible and intangible innovations, such as in household projects context. There are multiple conceptualizations for intangible innovations, which indicate the importance of conceptual clarity to distinguish between tangible and intangible innovations. We attempt to develop required conceptual clarity that will extend academic knowledge on understanding the phenomena, the nature of it as an individual project instead of idea, object or practice, its diffusion, and the impact of it on a national scale. Therefore, the exploration of studying this phenomenon will have theoretical and empirical implication as academic research.

The application of big data analytics approach can be utilized (e.g. they are growing in other sectors (e.g. in higher education [44]). The findings of this study in the future would be used to develop innovative ICT application design for end users following design guidelines that are established in other associative areas of functionalities for empowering end users [2, 45, 46] and highlighting features of building end-user specific service-based systems [2, 47, 48].